# Corporate Competition: A Self-Organized Network

Dan Braha

New England Complex Systems Institute, Cambridge, Massachusetts and University of Massachusetts, Dartmouth, Massachusetts

Blake Stacey and Yaneer Bar-Yam

New England Complex Systems Institute, Cambridge, Massachusetts

A substantial number of studies have extended the work on universal properties in physical systems to complex networks in social, biological, and technological systems. In this paper, we present a complex networks perspective on interfirm organizational networks by mapping, analyzing and modeling the spatial structure of a large interfirm competition network across a variety of sectors and industries within the United States. We propose two micro-dynamic models that are able to reproduce empirically observed characteristics of competition networks as a natural outcome of a minimal set of general mechanisms governing the formation of competition networks. Both models, which utilize different approaches yet apply common principles to network formation give comparable results. There is an asymmetry between companies that are considered competitors, and companies that consider others as their competitors. All companies only consider a small number of other companies as competitors; however, there are a few companies that are considered as competitors by many others. Geographically, the density of corporate headquarters strongly correlates with local population density, and the probability two firms are competitors declines with geographic distance. We construct these properties by growing a corporate network with competitive links using random incorporations modulated by population density and geographic distance. Our new analysis, methodology and empirical results are relevant to various phenomena of social and market behavior, and have implications to research fields such as economic geography, economic sociology, and regional economic development.

# I. INTRODUCTION

In recent years, major advances have been made in understanding the structure and dynamics of real-world social, biological, and technological complex networks [1–4]. Complex networks theory has also contributed to organizational and managerial environments, where new theoretical approaches and useful insights from application to real data have been obtained [5–7]. Through theory and experiment, these studies have characterized the structural properties of such networks, their mechanisms of formation, and the way these underlying structural properties provide direct information about the characteristics of network dynamics and function. Of particular interest are scale-free networks where the degree (i.e., the number of nodes adjacent to a node) is distributed according to a power law or a long right tail distribution. Such networks have characteristic structural features like "hubs", highly connected nodes [8], features which cause them to exhibit super-robustness against failures [9, 10] on the one hand, and super-vulnerability to deliberate attacks and epidemic spreading [11] on the other. Modeling real world large interfirm competition networks, which capture the coupling between economic units, is important to understanding the complex dynamics, robustness, and fragility of economic activity.

Here, we use network methodology to analyze and model the spatial structure of a large competition network, representing competitive interactions among firms within the United States. We find that the framework of geographic complex networks, mainly applied to natural and engineered systems, can be extended to capture the underlying structure and micro dynamics of interfirm competition, a system of heterogeneous economic units involved in strategic interaction. We extend the understanding of organizational problems by following a complex systems approach [1, 3, 5, 6, 47, 48]. To study complex systems, comprising many interacting units, we first look for robust empirical laws (often guided by prior knowledge) that describe the complex interaction followed by theoretical models that help understand and reproduce the main properties of the real world system. Our study combines several empirical measurements of competition networks and theoretical models, which are then validated and informed by the actual measurements. We focus initially on four fundamental properties: node degree distributions, the spatial distribution of firms, the relationship between connectivity probability and geographic distance, and edge length distribution. We then propose two simple models where new firms are added to the 2D surface of the Earth, and become connected to existing firms following cumulative advantage rules that are also dependent on geographic distance. We show that the models are able to reproduce the observed measurements remarkably well.

We represent corporate entities by network nodes, and we connect nodes using directed edges following the competitive relationships of the firms involved. The competitive ties between pairs of firms can be defined in two ways: The first approach is based on defining competition ties in terms of the potential for two firms to compete. According to this view, two companies are rivals if they are involved in direct or indirect competition for customers that have brand preferences for products (or services) that perform the same function, or products (or services) that are close substitutes for each other. The potential for two firms to compete is also defined in terms of the degree of intersection or overlap of the set of resources and environmental conditions that sustain the functioning of firms [52–55], such as financing on the capital markets. Following this approach, firms may compete even though they lack awareness of one another's existence and therefore cannot take one another's actions into account [52]. The second approach to competition, and the one we mostly use in this paper, views competition as a property of observable and consciously recognized social ties by the firms about the relation [47]. According to this approach, competitive ties among pairs of firms are based on the perspective, perceptions and evaluations held by the firms about the competition relations. Understanding the structure of the entire network of perceived competitive ties among firms is valuable for several reasons. It has been observed that participants in the market know each other; and, moreover, in formulating their competitive strategies (e.g., price cutting) participants monitor and take into account the actions and intentions of only a limited number of perceived others [64, 66--69]. The perceived competitive ties can be measured in a variety of ways (also used in this paper), including archival records (e.g., public filings, annual reports, and newspapers), interviews, or speeches.

While most network-theory research has concerned nodes and connections without any reference to an underlying space, in many real-world networks nodes and links are embedded in a physical space. In such networks, the interactions between the nodes depend on the geometrical distance between nodes; often, edges tend to link nodes that are close neighbors. Examples include natural, engineered, and social networks [12] such as the physical arrangement of the Internet [13–16], road and airline networks [16–18], broadcast signaling networks [19], power grids [20], mobile communication networks [21], and neuronal networks [22]. In real-world systems, the probability that two nodes are connected has been seen to decrease as a power law [14, 21, 23] or an exponential [15] of the distance between them. Other research has characterized the geographical deployment of nodes in two or three-dimensional Euclidean space. For example, Yook et al. [14] and Lakhina et al. [15] have shown that in technologically developed countries the Internet demand (measured by router density) is proportional to the population density. Several models of spatial networks have been proposed in the complex systems literature among which are placing nodes on simple regular lattices that are either joined randomly depending on their distance or if their distance is less than a certain cutoff [24, 25]; combining network growth and preferential attachment modulated by distance selection mechanisms [14, 26, 27]; and generating geographic networks based on local optimization processes [16].

Concerns of geographic and social proximity are not unknown in the social sciences. In sociology, gravity-based models predict that the likelihood of a relationship is inversely proportional to the physical distance between two individuals [43, 44]. In the context of international economics, the gravity model of trade predicts trade-flow volumes and capital flows between two units to be directly proportional to the economic sizes of the units (using GDP data) and inversely proportional to the distance between them [45]. In economic geography, the gravity model was used to explain migration flows between countries, regions, or cities [46], and showed that movement of people between cities is proportional to the product of their population size and inversely proportional to the square of the distance between them.

Spatial networks have also been of interest to economic geographers, who have considered networks as a means for understanding urban growth, geographical clusters, international trade, and globalization [28]. These efforts, however, have been mostly metaphorical and insufficiently formalized [29]. In sociology and organization theory, models of networks (including spatial networks) have largely focused on the factors that affect the dynamics of the formation of linkages between members of a network [30–34]. These empirical studies provide support for preferential attachment type of mechanisms [8] as an important driver of tie selection [34–37]. For example, the alliance behavior of multinational corporations indicates that firms will be more likely to have further alliances in the future with increasing experience and connectivity [33, 36], and an expanding network of interfirm alliances in American biotech exhibits preferential attachment [37]. Geography as a significant determinant of tie selection and network expansion has also been demonstrated. Empirical research illustrates that ties between firms, representing alliances, corporate board interlocks, or investments, are more likely when two firms are co-located [34, 37–39]. Moreover, studies show that geographical proximity affects the entry of firms in a network forcing them to locate in spatial proximity to industry agglomeration [40, 41].

In Section 2, we represent real-world data on corporate competition and headquarter location as a directed network in space. In Section 3, we report an asymmetry between the out-degree (number of corporations a firm is affected by) and in-degree (number of corporations a firm affects) distributions. Next, in Section 4, we report that the geographic arrangement of corporate headquarters strongly correlates with population density and that the probability two firms are competitors declines with geographic distance. In Section 5, we develop two models for spatial network growth that yield both the degree distributions and geographic statistics of the empirical network. We present empirical evidence and a theoretical scaling argument showing the close relation between the two models. We conclude in Section 6 with implications for the field of economic sociology. The Supplementary Material includes comprehensive statistical analyses of the various empirical data and models.

## II. CORPORATE GEOGRAPHIC COMPETITION NETWORK DATA

The competition network was reconstructed from information records provided by Hoover's – a large business research company that offers comprehensive business information through the Internet on corporations and organizations in over 600 industries. Within the detailed company records, information can be found on location type (headquarters or other); street, city and state address; financial information; industry codes; and competitors list. The competitors list was selected based on various information sources including public documents (e.g., SEC filings), company websites, industry-specific trade publications and journals, and directly from the company themselves. For example, Google lists 10 competitors by name in its filed 2009 10-K report to the U.S. Securities and Exchange Commission, including: 1) Traditional search engines, such as Yahoo! Inc. and Microsoft Corporation's Bing; 2) Social network sites, such as Facebook, Yelp, or Twitter; 3) Vertical search engines and e-commerce sites, such as WebMD (for health queries), Kayak (travel queries), Monster.com (job queries), and Amazon.com and eBay (commerce). While Google lists specific e-commerce and social network sites, Yahoo! Inc. – one of the leading U.S. search site with a broad range of other services – lists fewer competitors on its 2009 10-K report, including Google, Microsoft, AOL, Facebook and MySpace. Microsoft identifies in its 2009 10-K report competitors for each of its five operating segments: Client, Server & Tools, Online Services, Business, and Entertainment & Devices. In Appendix A, we provide a sample of Microsoft's products and services, competitors, and the rationale for competition, as identified by Microsoft.

In order to avoid problems of disjoint maps, we limit our study to firms with headquarters locations in the contiguous United States, for which detailed information on competitors was available. A large firm can also have many local or regional offices where the firm's activities are conducted. For example, Google is headquartered in Mountain View, California, but has branches in other US cities including Atlanta, Boston, Chicago, New York and Washington, D.C. [42]; a major industrial manufacturer can have its main corporate offices in one city and factories scattered elsewhere. In this paper, however, local and regional offices are not included in the competition network, because detailed and complete information regarding their list of competitors was not available. Still, the focus on headquarters location provides useful and direct information about the characteristics of competition; headquarters regularly gather data and intelligence from other competitors, and use the material collected to generate solutions to complex problems and identify competitive strategies [65]. Also, headquarters depend regularly on overlapping resources (e.g., workforce, investment banks, advertising and media companies, and consulting firms) that tend to cluster near one another. Finally, we will see in Section 4 and 5 that headquarters location is a meaningful quantity with considerable predictive power.

The competition network can be studied by several sampling methods [50]. Here, we use "snowball sampling" (e.g., [51]) starting from a single node (company), we select all of the nodes directly linked from it, then the nodes linked from those selected in the last step, continuing until a termination criterion is reached. To implement the method, we have created a Web crawler that browses the Hoover's website, and automatically downloads relevant information for subsequent analysis. The sampled competition network was obtained after crawling the web for a period of time that generated at least 10000 nodes -- a notably large network. Snowball sampling is a useful technique when relational data is not given explicitly. In our case, Hoover's maintains the corporate data in a website with the following structure: webpages correspond to different companies, and each webpage includes general data related to the company as well as a list of companies judged to be competitors. Thus, to construct the network, we perform snowball sampling. We begin with a company and collect its list of competitors; in a recursive fashion we traverse the list of competitors to other webpages, collecting more list of competitors and so on. This method could possibly generate a network that does not reflect (in a statistical fashion) the structure of the "real" network, because we start the sampling from a particular node. To eliminate this bias, one can construct different networks by starting the same snowball sampling method from different seed companies. After one computes several sampled networks $G_1$, $G_2$, $G_3$, ..., one takes the union of these networks (the set of nodes is the union of the set of nodes in $G_1$, $G_2$, $G_3$, ..., and duplicated arcs are excluded) to get a larger sampled network. The original sampled networks include companies around the globe, while the combined network analyzed is reduced to companies that operate within the United States.

Our sampling of the business information site, combining sampled networks starting from companies whose main activities are in different industries, resulted in a directed network of 10753 companies and 94953 competition links. The average in-degree (or out-degree) of a node is about 9, with 40% of the competition links being reciprocal. We analyze the structure of the strong components of the corporate competition network, and find that it has 420 components: one giant component that includes 10234 nodes; and 419 smaller components with varying sizes between 1 and 8.

Several observations indicate that one should study the interfirm competition network as an aggregate, rather than subdividing it by industry or company size. First, many competition relationships cross industrial boundaries. We use the North American Industry Classification System (NAICS), employed by various organizations to categorize businesses in Mexico, the United States and Canada, to sort the nodes in our network by economic sector. NAICS codes provide a hierarchical classification: the first two digits indicate general sector membership (22 meaning "Utilities" and 53 meaning "Real Estate, Rental and Leasing", for example), and later digits specify economic roles in more finely-grained detail. At the broadest level of classification, the first two NAICS digits, only 74% of interfirm competition links connect firms in the same sector. Focusing to the industry level, when we use the first five NAICS digits, this figure drops to 46%. If we use all six digits of the NAICS system, only 38% of links connect nodes of the same classification. Thus, while competition does appear to be somewhat sorted by industry (as is only natural), neglecting competition relationships which cross industrial boundaries would discard important properties of the system. A similar concern applies to corporate size: if we measure the size of a firm by its number of employees, the size of a company correlates only weakly with the average size of its neighbors in the network. (The Spearman's non-parametric correlation coefficient is only 0.56.) Therefore, subdividing the network by corporate size would also be an artificial and unilluminating choice.

### III. ANALYSIS OF IN- AND OUT- DEGREE DISTRIBUTIONS

A competition network can be considered as a directed graph with $N$ nodes and $L$ arcs, where there is an incoming arc to company $v_i$ from company $v_j$ if company $v_i$ lists $v_j$ as a competitor.

We compared (see Figure 1) the cumulative probability distributions $P_{in}(k)$ and $P_{out}(k)$ that a company has more than $k$ incoming and outgoing links, respectively. Several common parametric statistical distributions have been considered, and the fit of the different distribution models have been compared using likelihood ratio methods (see Appendices B and C for details). Both the out-degree and in-degree distributions can be described by a stretched exponential function [57, 59] of the form $x^{\beta-1}e^{-\lambda x^{\beta}}$. However, the in-degree distribution decays faster than the out-degree distribution ($\beta_{in} = 1.43, \beta_{out} = 0.53$), implying that companies with large in-degree are practically absent. This can also be seen as the markedly curved-shaped behavior of the in-degree distribution as shown in Figure 1. A possible explanation is that while companies typically consider as competitors a small number of other companies, there are a few companies that are considered as competitors by many others. This finding is supported by studies of competition, which demonstrate that firms formulate their competitive strategies by taking into account the perceived action only of a limited number of others [e.g., 64, 66 -- 69]. One could also think that the asymmetry between the in- and out- degree distributions is due to an upper bound on the number of competitors a firm could list. Another explanation is the limited amounts of resources and capabilities allocated for competition, which may limit the number of competitors a company considers. Independent information sources used for data collection -- including public documents, company websites, industry-specific trade and journals, and interviews -- suggest that the asymmetry could be attributed to a variety of limitations (whether cognitive or non-cognitive) including those mentioned above, network (10753 nodes, 94953 arcs). The out-degree distribution is characterized by a stretched exponential law $x^{\beta-1}e^{-\lambda x^{\beta}}$ with $\beta = 0.53$ and $\lambda = 0.58$ (see Appendices B and C). The in-degree distribution is fitted by a stretched exponential with $\beta = 1.43$ and $\lambda = 0.04$ (see Appendices B and C). (B) Several sampled networks. The networks were identified by snowball sampling, starting from a seed company whose main activity is in a particular industry (Energy & Utilities: 4248 nodes, 35789 arcs; Transportation Services: 3408 nodes, 31330 arcs; Computer Hardware: 6914 nodes, 56663 arcs; Construction Industry: 4094 nodes, 35376 arcs; Agriculture: 3245 nodes, 32758 arcs). We find that the out-degree distributions of all sampled networks are similar to each other as well as to the aggregate network, suggesting the robustness of the snowball sampling used here.

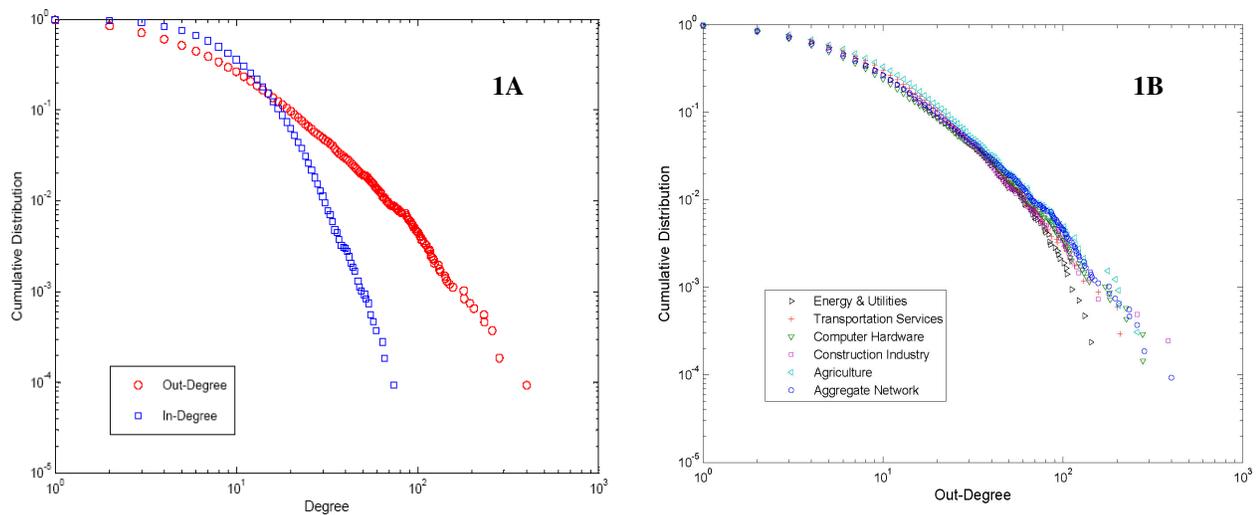

FIG. 1: The log-log plots of the cumulative distributions of incoming and outgoing links. (A) Aggregate competition

Asymmetric in- and out- degree distributions have been found in other large complex networks [5, 48]. The connectivity of competition networks is important in constraining and determining many aspects of dynamical processes occurring on top of them, such as pricing decisions, strategic behavior, and firm performance. For example, it stands to reason that events and activities of central firms will tend to quickly propagate (due to the heterogeneous outgoing connectivity) throughout the entire competition network, benefiting or impairing the vitality of the interconnected firms. This seems similar to ecological networks, where the loss of a keystone species could have large effects on the network [52].

We next examine the spatial characteristics of competition networks. The specific latitude and longitude of each company was obtained from its address using Yahoo's Geocoding Web Service, and the distance between two companies was calculated by using their geographical coordinates. In Figure 2, we compare the geographical deployment of companies with the population distribution in the contiguous U.S. The high correlation found between the spatial deployment of companies and population density is intuitive and may not be surprising [74]. Indeed, the patterns shown in Figure 2 can be explained as follows [65]. Firms depend and interact regularly with business service providers. Both firms and service providers (e.g., investment and commercial banks, and consulting firms) tend to cluster in areas that can attract and retain a highly skilled workforce; have a large market to their products; and have convenient access to airports, highways, and telecommunication infrastructure. This implies that firms, their competitors, and their business service providers tend to be located near one another, with a bias towards large and densely populated metro areas. Here, however, we attempt to go beyond the simple correlation shown in Figure 2, and to highlight a similarity (rather than simple correlation) between the actual spatial distribution of companies and the actual population density. The similarity between the two different spatial distributions was explicitly utilized in the competition models presented in Section V. More specifically, in the competition models, firms are distributed on the Earth's surface by sampling from the population density distribution. From a dynamic point of view, this also points to possible common causal mechanisms that couple the processes of population dynamics and new firm emergence.

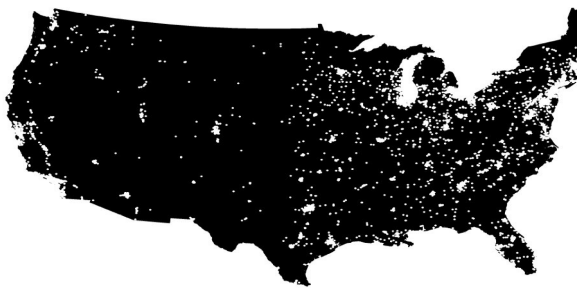 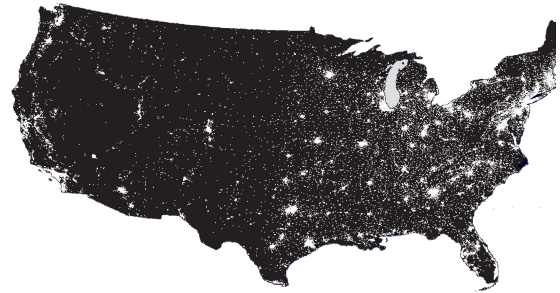

**2A. Geographical Deployment of Firms**  **2B. The 2000 Population Distribution in the Contiguous United States**

FIG. 2: (A) Geographical deployment of companies with headquarters locations in the contiguous United States. The latitude and longitude of each company was obtained from its address using Yahoo's Geocoding Web Service. (B) Map of the 2000 Population Distribution in the United States (also referred to as the "Nighttime Map") produced by the U. S. Census Bureau as part of the 2000 Decennial Census. In this map, white dots coalesce to form the urban population concentrations; each white "dot" represents 7,500 people.

## IV. ANALYSIS OF GEOGRAPHICAL DISTRIBUTIONS

The corporate competition spatial network enables us to relate competition and geographic distance. Figure 3 shows the probability $P(v_i \rightarrow v_j | d(v_i, v_j) = l)$ that two companies separated by a distance *l* are related by a competition link, indicating that geographic proximity tends to increase the probability of competition. In Appendix D, we consider several alternative models for the probability of competition, and provide detailed information about parameter estimation and statistical model comparison. The results show that as the distance *l* increases, the decrease in competition probability can be plausibly described as a power-law. However, the fluctuations around the power-law behavior for distances larger than ≈ 1000 km also imply that a model for the presence of competition needs to take into account both geography-dependent mechanisms and non-geographic processes, as explored later in this paper.

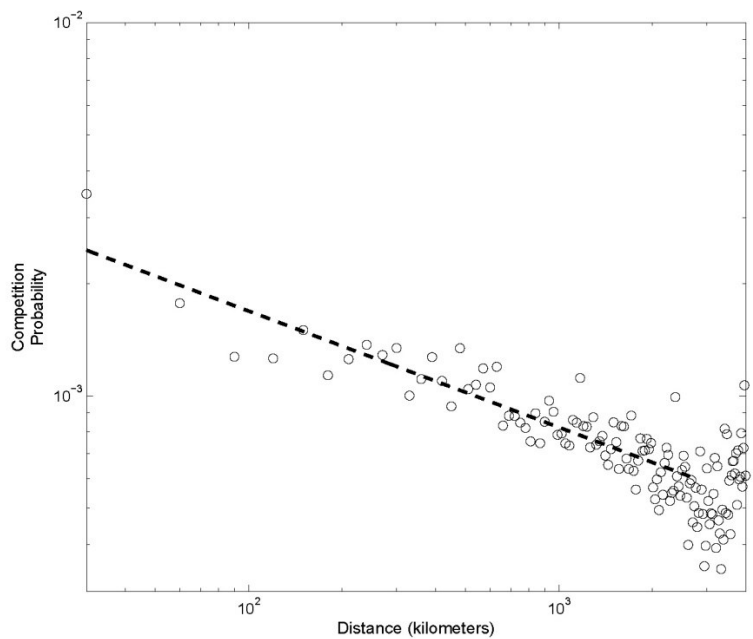

FIG. 3: The relationship between geographic distance and competition. The log-log plot shows the probability that two companies separated by a distance l are related by a competition link. The competition probability is fitted by a power-law $l^{-\beta}$ with β ≈ 0.3 (see Appendix D). The probability was estimated from the proportion of pairs of connected companies separated by a distance *l* among all the total number of pairs of (connected or not) companies separated by a distance *l* (practically, link lengths with a resolution of 30 km are examined).

The physical distance between nodes in geographic networks plays an important role in determining the costs and benefits of communication and transport. As such, common to many geographic networks is a bias towards shorter links. The competition network analyzed here is of no exception (but perhaps for different reasons). Indeed, we see in Figure 4B that the competition network has many very short links of length 100 km or less. However, the competition network also includes a large portion of links of length 3800 km or less, and then an apparent smaller peak of longer links around 4000 km. Many of these longer links represent continent-wide distances. We show in Figure 4A the cumulative probability distribution that the length of a link is greater than $l$ kilometers. In Appendices B and C we analyze several common parametric statistical distributions, and find that the link length distribution is best fitted by a power law with subsequent exponential decay of the form $l^{-\gamma}e^{-\lambda l}$. The estimated parameter $\lambda = 0.0005$ indicates that the characteristic distance beyond which the probability distribution is dominated by an exponential decay is about $1/\lambda \approx 2000\,\text{km}$.

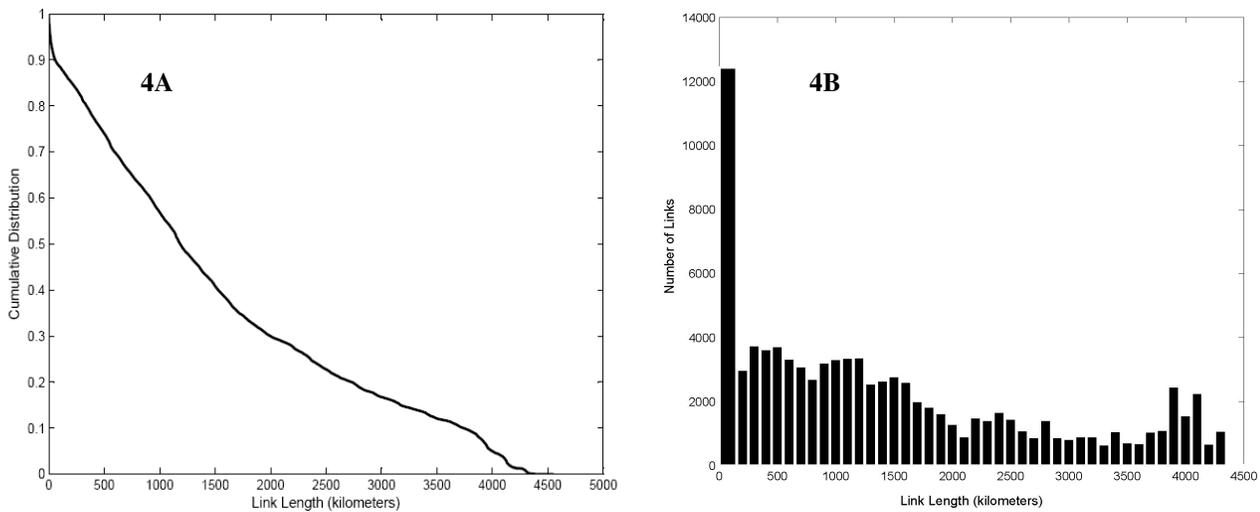

FIG. 4: (A) The cumulative probability distribution that the length of a link is greater than $l$ kilometers. The cumulative distribution is fitted by a power law with exponential cutoff $x^{-\gamma}e^{-\lambda x}$ with $\gamma = 0.17$ and $\lambda = 0.0005$. (B) The histogram of the lengths of links. We see that the competition network has many very short links of length 100 km or less, a large portion of links of length 3800 km or less, and then an apparent smaller peak of longer links around 4000 km. Many of these longer links represent continent-wide distances.

The geographic nature of the competition network also has an effect on its topological robustness. In network theory, "robustness" refers to a network's ability to withstand attacks, such as random deletion of nodes or the targeted removal of highly-connected hubs. The effect of attacks is typically gauged by the change effected in the network's topological properties, such as the size of its largest component: a network which falls apart into many small pieces upon the excision of a single node is fragile. Power-law networks grown by preferential attachment have been found to be resilient against random attacks, but weak against the targeted deletion of high-degree nodes. Here, we see that the degree distribution is not the only relevant factor in determining robustness; as Figure 5 shows, even after we delete thousands of nodes, the network does not dissolve into disconnected pieces.

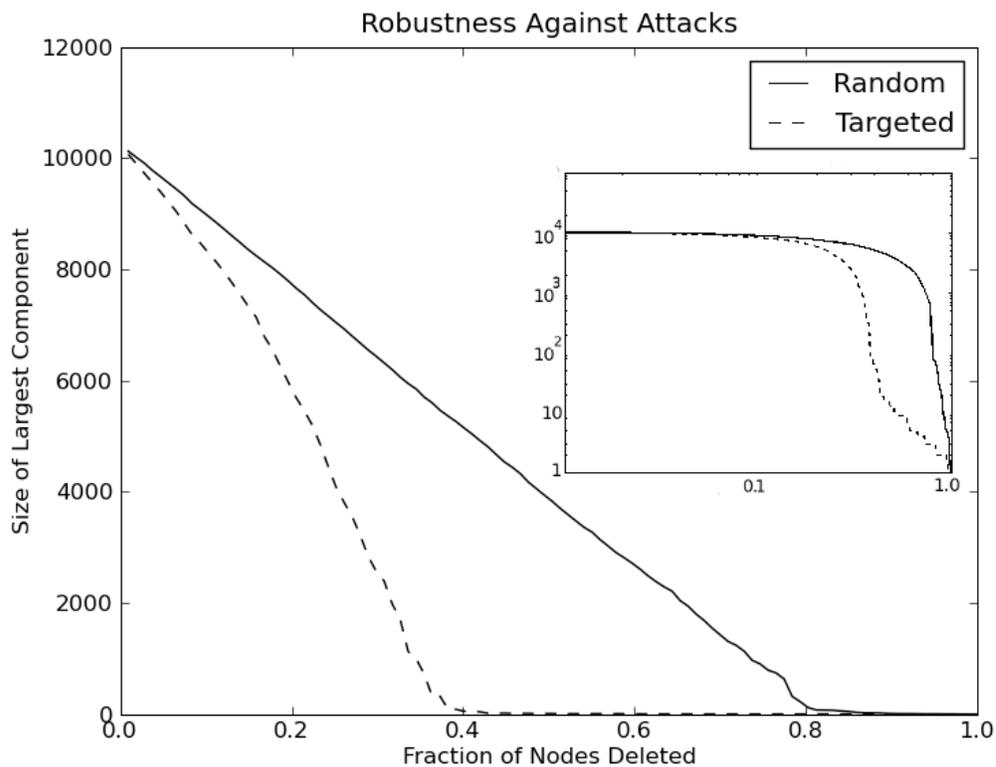

FIG. 5: Robustness of the competition network, as demonstrated by the deletion of nodes. Solid (upper) line shows the size of the largest strong component surviving as nodes are randomly deleted; Dashed (lower) line shows the size of the largest component as nodes are deleted in order of decreasing out-degree. Higher-degree nodes are more "central" in that attacking them breaks apart the network more efficiently. Inset: the same data, plotted logarithmically.

## V. MODELING CORPORATE SPATIAL COMPETITION NETWORKS

Motivated by the above empirical results, we develop in this section two explanatory mechanisms that provide insight on the organization of interfirm spatial competition networks. The two mechanisms are classified into two types. The first model incorporates a geography-dependent mechanism combined with a "preferential attachment" type of rule [8], where the topology of the competition network evolves in such a way that already well connected companies have greater probability to attract more competition links in a process of positive feedback (the "the rich get richer" paradigm). The first model is particularly appropriate for modeling network formation whose future topology is largely determined by past events that are "internal" to the network. Thus, we also consider a second model where the future topology of the network is determined by "external" information, which is non-topological in nature, and is intrinsically associated with each company in the network. Specifically, the second model incorporates a geography-dependent mechanism combined with a "fitness" type of rule, where companies are assigned an intrinsic quantity or fitness (which *could* change in time) in such a way that the probability of attracting more competition links is related to the fitness of the company (similar concepts of "fitness" have been used in various studies including "firm size" [69], "talent" and "quality" [70], or "status" [35]). We show that both types of models are able to reproduce empirically observed characteristics of competition networks. More importantly, despite the seemingly different premises of the two models, we provide a scaling argument showing the equivalence of the two models. The scaling argument is supported by the empirical results and numerical simulations.

**V.1. Preferential Attachment and Geography-Dependent Mechanism**

Although the competition network studied here represents data collected at particular time point, it is the result of a specific development path of network dynamics that involve firm entries and exit as well as the formation and dissolution of competition links. The heavy tail characteristic displayed by the out-degree distribution of the competition network (Figure 1) points to the possibility that the formation of competition networks could be governed by a preferential attachment rule [8]. Figures 3 and 4 show that firms compete according to the distance between them, signifying that the formation of links is also geometrical in nature. Moreover, the strong correlation between firm and population distributions (Figure 2) means that any model of competition should take into account the concentration of firms in highly populated areas.

The preferential attachment and geography-dependent mechanisms can be explained using substantive concepts that pertain to the nature of competition. In economics, "barriers to entry" refer to circumstances that make it very difficult or costly for a potential new entrant to compete with established firms that are already selling competing goods or services. Accordingly, incumbent firms with few competitors ("lower degree" as in a monopoly) have high entry barriers; and incumbent firms with lots of competitors ("higher degree" as in perfect competition) have low entry barriers. The preferential attachment mechanism -- where the more competitors a firm has (and thus the lower the entry barrier), the more likely it is to receive new competition links -- is a dynamic manifestation of the barrier-to-entry effect. The preferential attachment rule can also be explained in terms of "spin-out" companies of established firms. Accordingly, it is more likely that incumbent firms with many competitors will attract more competition links from "spin-out" companies of the direct competitors. The geography-dependent mechanisms can be explained in terms of the benefits for firms to cluster near one another. One of these benefits is the knowledge spillover that occurs among firms (within and between industries), which fosters the exchange of ideas and rapid adoption of innovation.

We therefore seek a model that considers the interplay between preferential attachment, geographic distance, and population density effects. A real understanding and modeling of competition networks should be able to reproduce empirically observed characteristics of competition networks — such as the degree and link length distributions reported in Figures 1 and 4 — as natural outcome of a minimal set of general mechanisms governing the formation of competition networks.

Network growth models including geographical distance of nodes [14, 25, 26] are a natural modeling approach for competition networks. We identify points on the curved surface of the Earth by their latitude and longitude coordinates, and compute geographic distances using the great circle distance between pairs of points on the surface of a sphere. We then superimpose on the map a grid consisting of two sets of parallel longitude and latitude lines, dividing the Earth's surface into squares (for our numerical simulations, we use high resolution data that consists of boxes of $0.0083° \times 0.0083°$). At each box, the population density is calculated from population data by dividing the population of each box by its area in square kilometers. In the following, firms are distributed on the Earth's surface by sampling from the population density distribution. We start with $m_0$ firms, each pair connected by a competition link, and at each subsequent step the network grows with the addition of new firms. For each new firm, $m$ new directed competition links are created connecting it to firms already present in the system. The exponentially truncated power-law distribution of the outgoing connections (Figure 1) suggests the use of a nonlinear preferential attachment rule [1, 3], which generalizes the linear preferential attachment mechanism that results in a power-law degree distribution [8]. This, combined with the fact that the competition probability tends to decrease with geographic distance according to a power-law, offers the possibility that the growth of competition networks is governed by a nonlinear preferential attachment rule modulated by a link length dependent factor. More specifically, the firms $j$ receiving the new links from firm $i$ are chosen with probability proportional to $k_j^{\alpha} l_{ij}^{\beta}$, where $k_j$ is the total degree of firm $j$; $l_{ij}$ is the length in kilometers of the directed link from $i$ to $j$; $\alpha$ and $\beta$ are continuously varying parameters.

We have tested the validity of the above model by conducting several extensive computational experiments, and comparing the simulation results with the actual observations made from the competition network. In the simulations below, we have used a population density grid from the 2000 U.S. Census produced by the Columbia University Center for International Earth Information Network (CIESIN). The grid has a resolution of 30 arcseconds (0.0083 decimal degrees), or approximately 1 square km. In all cases, we start with $m_0 = 8$ connected firms, and at each step of network growth a new firm with $m = 8$ directed competition links will be connected to firms already present in the system until the total number of firms reaches the actual number of firms in the sampled competition network $N = 10753$.

The model above offers a good flexibility for calibration since two parameters can be modified. Altering the value of α and β will influence the estimated spatial interactions. As described in Appendix E, we have calibrated our model to correctly reproduce the empirically observed degree and link length distributions characteristics reported in Figures 1 and 4, and have derived the values of $\alpha = 0.85$ and $\beta = -0.3$ (henceforth called Competition model 1). The plausibility of the proposed model given the observed data is assessed by comparing the proposed model with a null model, which serves as the baseline for making comparisons of improvement in model fit. In our case, the null model corresponds to the parameters $\alpha = 0$, $\beta = 0$, and where firms are distributed randomly and uniformly on the surface of the contiguous U.S. That is, the null model maintains the growing character of the network, but the "preferential attachment" and "friction of distance" are eliminated by assuming that a new firm is connected with equal probability to any firm in the system. In addition, we examine three extreme cases of the competition network model: (1) Linear Preferential Attachment: $\alpha = 1$, $\beta = 0$; (2) Gravity I: $\alpha = 0$, $\beta = -1$; and (3) Gravity II: $\alpha = 0$, $\beta = -2$. The first case corresponds to the scale-free network model developed by Barabasi and Albert [8] where an already present firm receives a new competition link a according to a linear preferential attachment rule, that is, with probability proportional to its degree. The second and third cases reflect a variety of gravity models in social science that are based on the empirical principle that proximity in geographic (and social) space affects the likelihood of interaction [43–46]. Validation and accuracy assessment of the various network growth models is performed by visual comparisons (see Figures 6-9 below), as well as quantitative analysis as detailed in Appendix E.

Figures 6A and 6B compare the link length and out-degree distributions, respectively, generated by the above five models with the empirically observed distributions shown in Figures 1A and 4A. We note that both the link length and out-degree distributions of the real competition network deviates significantly from that produced by the Null, Linear Preferential Attachment, Gravity I, and Gravity II models. The simulation results, however, of Competition model 1 are able to nicely reproduce the actual observations of the competition network, indicating that Competition model 1 gives a better characterization of the data than the four models specified above. Competition model 1 takes into account three effects: population density, preferential attachment, and geographic distance. The value of $\alpha = 0.85$ reflects a sublinear ($\alpha < 1$) tendency of preferential linking to firms with many competition links, which can result in a truncated power-law degree distribution [2, 5], as indeed observed empirically (Figure 1A). The value of $\beta = -0.3$ shows that the "friction of distance," or how rapidly interaction decreases as distance increases, is relatively small compared to the Gravity-based models.

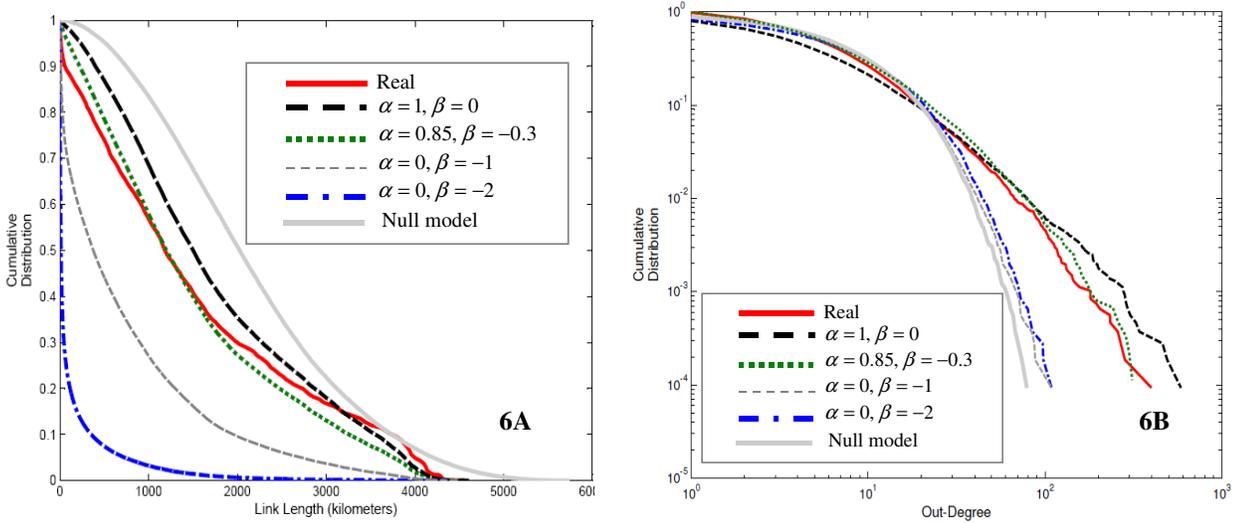

FIG. 6: Comparison between the actual competition network and simulation results of four network competition models, with respect to the cumulative link length (A) and out-degree (B) distributions. The models considered are: Linear Preferential Attachment (black dashed line, α = 1, β = 0); Competition model 1 (green dotted line, α = 0.85, β = −0.3); Gravity I (gray dashed line, α = 0, β = −1); Gravity II (blue dashed-dotted line, α = 0, β = −2); Null model (gray solid line, α = 0, β = 0). Distributions corresponding to the actual competition network are shown as red solid lines. In the simulations, we have used a 3120 by 7080 population density grid from the 2000 U.S. Census with a box resolution of 30 arc-seconds (0.0083 decimal degrees), or approximately 1 square km. In all cases, firm location on the surface of the contiguous United States is determined by randomly sampling from the population density distribution. Once a box is sampled, the firm's latitude and longitude are uniformly and randomly located within the box. (A) Comparison of cumulative link length distributions. (B) Comparison of cumulative out-degree distributions. The out-degree distribution generated by the Preferential Attachment model (with linear preferential attachment and without geographical distance effect) is fitted by a power-law; the out-degree distribution corresponding to Competition model 1 is characterized by an exponentially truncated power-law; and both Gravity I and Gravity II (with only geographical distance effect) generate distributions that are well fitted by an exponential.

In order to test for the effect of geographical distance bias on the competition network growth, we have held the "preferential attachment" parameter α at its optimal value 0.85 and have varied the "friction of distance" parameter β set at values 0, −0.3, −1, and −2. Figure 7A shows how a decrease in the value of β leads to a significant bias towards shorter links. However, Figure 7B shows that the out-degree distributions are almost not affected by the value of β, when α is set at the optimal value 0.85. This indicates that the parameter α has a strong effect on the out-degree distribution, and weak effect on the link length distribution.

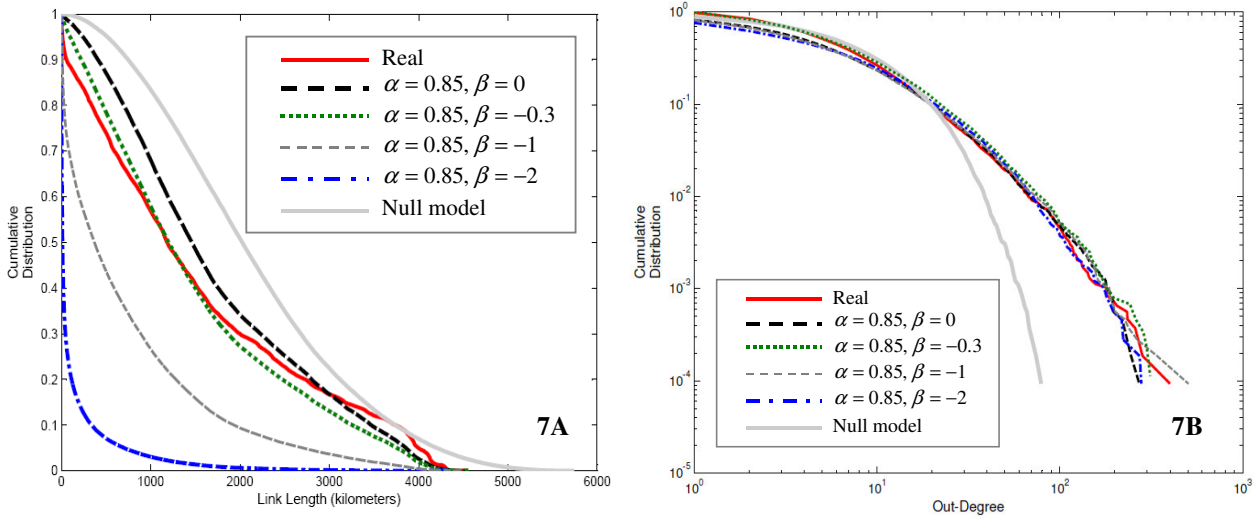

FIG. 7: The effect of the geographical distance bias on the link length (A) and out-degree (B) distributions. In the simulations, we have held the "preferential attachment" parameter α at its optimal value 0.85, firm location is determined by the population density distribution, and we have varied the "friction of distance" parameter β. Five network competition models are considered: Sublinear Preferential Attachment without Geographical Distance Bias (black dashed line, α = 0.85, β = 0); Competition model 1 (green dotted line, α = 0.85, β = −0.3); Sublinear Preferential Attachment with Inversely Linear Distance Bias (gray dashed line, α = 0.85, β = −1); Sublinear Preferential Attachment with Inversely Square Distance Bias (blue dashed-dotted line, α = 0.85, β = −2), and Null model (gray solid line, α = 0, β = 0). Distributions of the actual competition network are shown as red solid lines.

Testing for the effect of preferential attachment on the competition network growth further corroborates this finding. This is done by experimenting with varying values of α, when β is set at the optimal value −0.3. As shown in Figures 8A and 8B, while the link length distributions for varying α are not changed and are similar to the actual distribution, the out-degree distributions deviate significantly for values of α that are different from the optimal value 0.85. These results imply that the actual link length distribution is determined to a large extent by the "friction of distance" parameter β, and weakly so by the "preferential attachment" parameter α. Overall, Figures 7–8 show once more that Competition model 1 provides a better characterization of the competition network than other combinations of α and β.

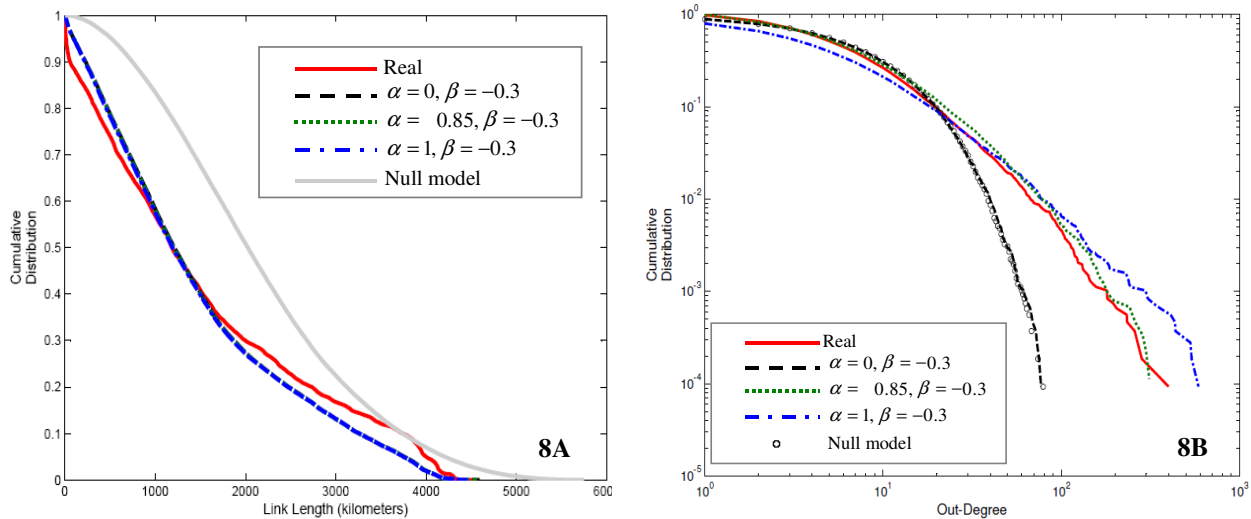

FIG. 8: The effect of the preferential attachment bias on the link length (A) and out-degree (B) distributions. In the simulations, we have held the "friction of distance" parameter β at its optimal value −0.3, firm location is determined by the population density distribution, and we have varied the "preferential attachment" parameter α. Four network competition models are considered: Geographical Distance Bias without Preferential Attachment (black dashed line, α = 0, β = −0.3); Competition model 1 (green dotted line, α = 0.85, β = −0.3); Linear Preferential Attachment with Distance Bias (blue dashed-dotted line, α = 1, β = −0.3); and Null model (gray solid line or black circles, α = 0, β = 0). Distributions of the actual competition network are shown as red solid lines.

Finally, we test for the effect of population density on the competition network growth. To this end, we have set the values of α and β to their optimal values, and have chosen the location of firms based on two methods: (1) Location by Population Density (as in Competition model 1): firms are distributed on the surface of the contiguous U.S. by sampling from the population density distribution; and (2) Random Location: firms are distributed randomly and uniformly on the surface of the contiguous U.S. Figure 9B shows that both firm placement schemes give similar results when comparing their out-degree distribution results with that of the observed data. However, as shown in Figure 9A, the link length distribution produced by the random location scheme deviates significantly from that produced by both Competition model 1 and the actual competition network. In summary, Figures 6–9 provide good evidence that the structure of competition networks can be better explained by taking into account network dynamical growth, preferential attachment, geographical distance, and demographic factors such as population density. In particular, a simple model that is able to reproduce reasonably well the main observed features was proposed.

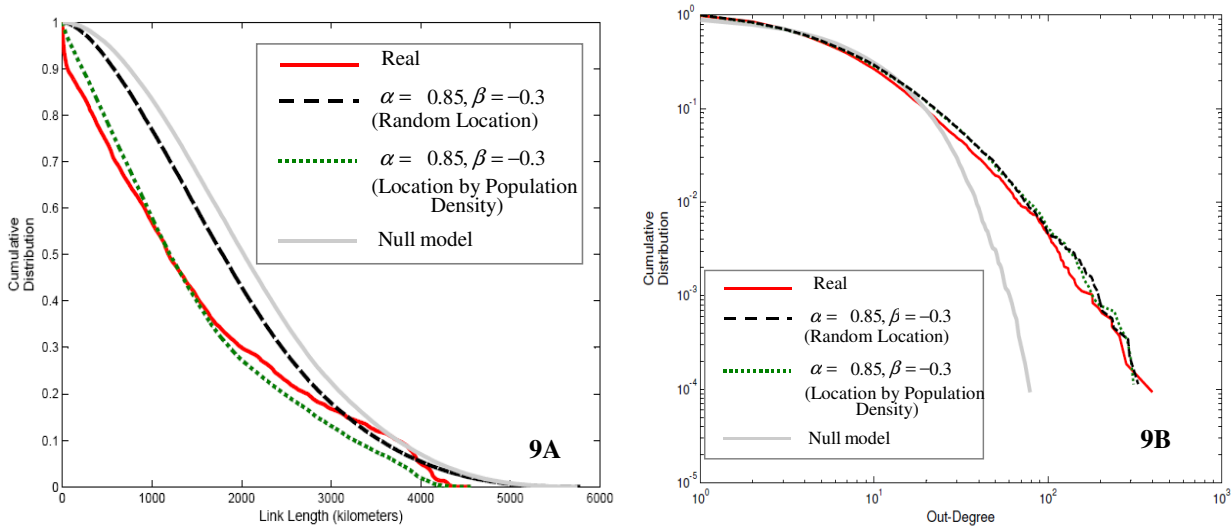

FIG. 9: The effect of firm location decisions on the link length (A) and out-degree (B) distributions. In the simulations, we have held the "preferential attachment" parameter α and "friction of distance" parameter β at their optimal values 0.85 and −0.3, respectively. Firm location is determined by two distinct mechanisms: Random Location (black dashed line) by which firms are distributed randomly and uniformly on the surface of the contiguous U.S.; and Location by Population Density (green dotted line) by which firms are distributed on the surface of the contiguous U.S. by sampling from the population density distribution. Distributions of the actual competition network are shown as red solid lines, and the Null model (α = 0, β = 0) is shown as gray solid line.

**V.2. Fitness and Geography-Dependent Mechanism**

While the first model presented in Section V.1 largely assumes that the competition network evolves according to an "internal" mechanism ("preferential attachment"), where the future topology is determined by past ones, it is possible that the topology of the competition network is determined by "external" information in the form of an intrinsic quantity or "fitness" associated with each company in the network. Here, we consider the size of a company in terms of the numbers of employees [69] as the natural candidate to be identified with the fitness associated with each company in the competition network (similar results are obtained when measuring size by sales, assets, or profits). In this section, we present an alternative competition model (Competition model 2) that incorporates directly the size of companies.

Before we present the model, we first present empirical evidence and theoretical scaling argument showing that the size of companies is scaled with their degree (number of competitors) in the competition network. To quantify the relationship between the size of companies and their degree in the competition network, we calculated the average size $\bar{s}$ of companies with out-degree $k$ in the competition network. For the competition network, the average size $\bar{s}$ shows a gradual increase with $k$, which can be fitted with a power law $\bar{s}(k) \propto k^\delta$ with $\delta \approx 1.07$ (Figure 10A). The fluctuations of the size of individual companies $s(k)$ around the mean $\bar{s}(k)$ have been found to be small; thus, we can make the following approximation: $s(k) \propto k^\delta$. If we approximate the out-degree distribution with a power law, with or without cutoff, $p(k) \propto k^{-\gamma}$, then

$$p(s) = p(k)\frac{dk}{ds} = \frac{p(k)}{s'(k)} \propto \frac{k^{-\gamma}}{\delta k^{\delta-1}} \quad (1)$$

Since $k \propto s^{1/\delta}$, we conclude that

$$p(s) \propto \frac{s^{-\gamma/\delta}}{s^{(\delta-1)/\delta}} \propto s^{-\gamma/\delta - (\delta-1)/\delta} = s^{-\sigma} \quad (2)$$

leading to the relation between the three scaling exponents:

$$\sigma = \frac{\gamma + \delta - 1}{\delta} \quad (3)$$

The significance of this result is that the right-skewed property characterized by the firm size distribution can be related to more fundamental scale invariant properties, characterized by the two exponents γ and δ. Equation (3) is confirmed for the competition network analyzed here. The calculation of $\gamma \approx 1.6223$ using maximum likelihood estimation (see Table B.2 in Appendix B) and equation (3) gives rise to $\sigma = 1.5816$, which agrees pretty nearly with the value of the exponent $\sigma \approx 1.5925$ as obtained by using maximum likelihood estimation to the firm size data (see Figure 10B). The heuristic utility of the scaling result can be seen by applying the scaling relation $k(s) \propto s^{1/\delta}$ to the attachment rule $k_j^\alpha l_{ij}^\beta$ used in Competition model 1 (see Section V.1); obtaining an attachment rule that incorporates directly the size of companies: $s_j^{\alpha/\delta} l_{ij}^\beta = s_j^\nu l_{ij}^\beta$, where $\nu = \frac{\alpha}{\delta}$. This leads to the possibility that "firm size" and "number of competitors" are two manifestations of *market competition signals*, which can thus be applied interchangeably for predictive purposes. The model presented below incorporates these insights.

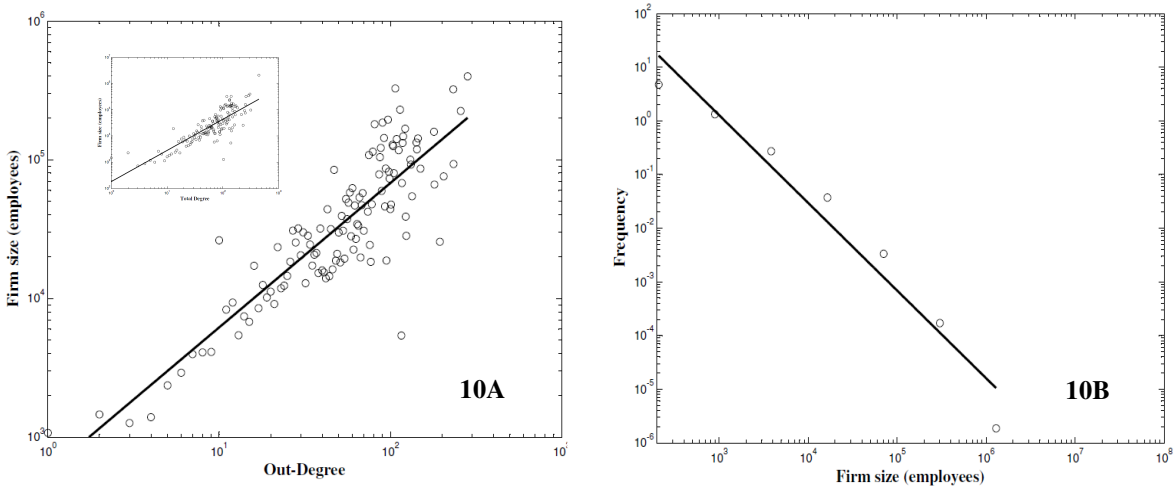

FIG. 10: Relationship between firm sizes, by employees, and their node degrees in the competition network. (A) The average size $\bar{s}$ of companies with out-degree k in the competition network. The solid line is a power law fit $\bar{s}(k) \propto k^{1.07}$. Inset: average size versus total degree shows a power law fit $\bar{s}(k) \propto k^{1.18}$. (B). Histogram of firm sizes, by 2009 employees, constructed using logarithmic binning having width increasing in powers of two. The exponent $\sigma \approx 1.5925$ of the power law was obtained using maximum likelihood estimation for values $s \geq s_{min}$ [57], where the minimum value $s_{min}$ was set to the lower end of the range shown in Figure 10A.

Motivated by the scaling argument presented above, we next present a model (Competition model 2) that takes into account explicitly the size of companies. We then show that Competition model 2 yields similar results to that reported for Competition model 1. Competition model 2 incorporates three mechanisms: (1) the geography-dependent mechanism is the same as discussed in Section V.1. That is, firms are distributed on the Earth's surface by sampling from the population density distribution, and the probability of attracting more competition links tends to decrease with geographic distance; (2) the probability of attracting more competition links increases with the size of the company; and (3) the firm size dynamics follows a Gibrat-like stochastic growth process in which growth rates are independent of size [73], and in which firm sizes are not allowed to fall below a lower bound that can change over time [71, 72]. As far as we know, there is limited effort in the network literature to connect models of "fitness" dynamics (in our case, firm size dynamics) with models of network formation (in our case, dynamics of linkage formation).

More specifically, we start with $m_0$ firms, each pair connected by a competition link, and at each subsequent step the network grows with the addition of new firms, which are assigned a size of "one" arbitrary unit. For each new firm, $m$ new directed competition links are created connecting it to firms already present in the system. The scaling relation between firm size and degree (Figure 10A) suggests the use of a nonlinear cumulative advantage rule, where the probability of attracting more competition links increases with the size of the company. More specifically, the probability of attaching to an existing firm $j$ is proportional to $s_j^\nu l_{ij}^\beta$, where $s_j$ is the current size of firm $j$; $l_{ij}$ is defined as in Competition model 1; and $\nu$ and $\beta$ are continuously varying parameters. Once the new node is attached, the sizes of already existing firms in the system are changed according to a Gibrat-like process with a reflective lower bound, known as the Kesten process [71, 72]. In particular, we use the model presented in [71], where at each step a firm is randomly selected and updated according to the following stochastic rule:

$$s_i(t+1) = \max\{\lambda(t)s_i(t), c\bar{s}(t)\} \qquad (4)$$

where $0 \leq c < 1$ is a constant factor, $\bar{s}(t)$ is the average size of firms at time $t$, and $\lambda(t)$ is a random growth rate drawn from a given density function $f(\lambda)$ that satisfies $\int_\lambda \lambda f(\lambda)\, d\lambda = 1$. Here we use, without loss of generality [71], a uniform distribution in the range $\lambda_{\min} \leq \lambda \leq$ $\lambda_{max}$. The stochastic growth process in Equation (4) is simulated for $T$ arbitrary units of time before a new firm is added to the competition network.

Figures 11A and 11B compare the link length and out-degree distributions, respectively, generated by Competition models 1 and 2 with the empirically observed degree and link length distributions shown in Figures 1A and 4A. Both models, which utilize different approaches yet apply common principles of "cumulative advantage" and "market competition signals," give comparable results that nicely reproduce the actual observations of the competition network.

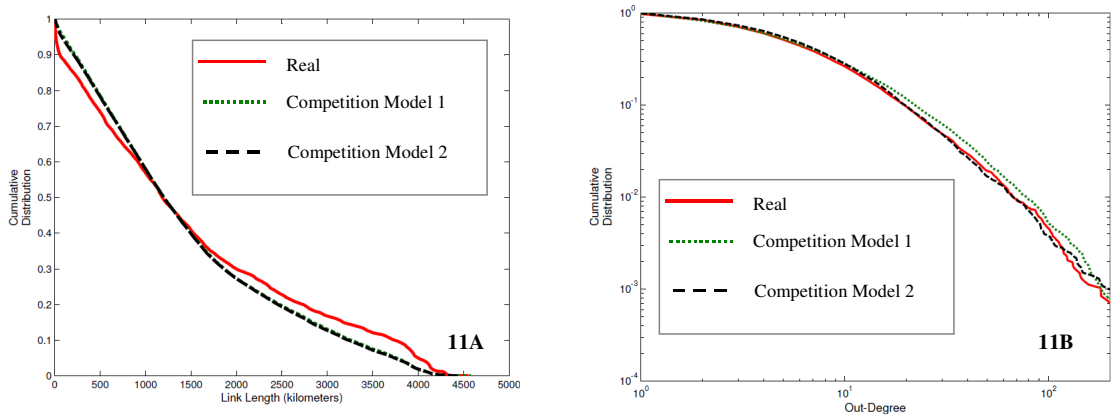

FIG. 11: Comparison between the actual competition network and simulation results of Competition models 1 and 2, with respect to the cumulative link length (A) and out-degree (B) distributions. We have used the following parameters for Competition model 2: the random growth rate λ(t) is set to be uniformly distributed between $\lambda_{min} = 0.9$ and $\lambda_{max} = 1.1$ [71], $T = 1000$, the "friction of distance" parameter $\beta = -0.3$ as in Competition model 1, and $\nu = 0.75$ and $c = 0.23$ were determined by fitting the parameters to the observed data.

## VI. CONCLUSIONS

We have analyzed a large inter-organizational network where the nodes are firms located in the U.S. and directed links represent competition by the nodes forming the link. We focused first on topological properties, and have shown that the competition network exhibits a noticeable asymmetry between the exponentially truncated power law distribution of outgoing competition links and the exponential law governing the in-degree distribution. This characteristic, which is consistent with results of other complex networks [5, 48], can be explained as follows: Firms are not regarded as passive economic entities, but the actions taken by firms could also be seen as determined by and affecting the behavior of other competitors. The exponential law governing the in-degree distribution could indicate a limitation on the firm's capacity to compete with (and thus affecting) many firms, while the power law governing the out-degree distribution could reflect the ability of competition networks to minimize the effects caused by major events or changes that require significant adjustment in firm behavior. Indeed, the power-law behavior of the out-degree distribution implies that there are only a few firms with many outgoing competition links (*i.e.,* affected by many others), which means that most of the time the competition network will display a low sensitivity to network perturbations. Altogether these results suggest that the structure of competition networks tend to stabilize the dynamics of competition.

The geographical aspect of the competition network has been analyzed in three ways. First, we have shown that the spatial distribution of companies is strongly correlated with the population distribution. This finding emphasizes the important role of environmental and exogenous mechanisms as context for network formation. Second, we have shown that geographic proximity increases the probability of competition following a power law, characterized by a scaling coefficient ("friction of distance") which is considerably lower than values used in other gravity-based models. This result could be explained, for instance, as a consequence of improvements in transport efficiency or communications technology, both of which tend to reduce the value of the friction of distance $\beta$. Third, we have analyzed the physical distance between firms, and have shown that the link length probability distribution is well fitted by a power law with exponential decay distribution with many very short links of length less or equal to 100 km, and extended link lengths of up to 4000 km. This is indicative of the tendency of competition networks to agglomerate into geographic concentrations ("clusters") of interconnected firms with characteristic size of about 100 km, and with competition links of varying lengths between separate clusters.

Motivated by the above empirical observations, we have proposed two models for the formation of competition networks, which are grouped in two classes according to whether the connection probability is determined by information that is internal ("degree") or external ("firm size") to the network. By comparing simulation results with the empirical observations of the competition network, we have demonstrated that both models are able to reproduce observed characteristics. The competition network models share two common features that provide insights into the factors governing the origin of competition networks: (1) spatial locations of firms, which are positively correlated with the population density; and (2) stochastic incremental growth governed by nonlinear cumulative advantage rule (determined by "degree" or "firm size") modulated by geographic distance. We have shown that the "degree" of a firm in the competition network (internal property) and its "size" (external property) can be treated on equal footing. This leads to the intriguing possibility that both "degree" and "firm size" are two instances of "market competition signals," by which a company conveys information to potential new competitors about the future possibility of success once competition is established.

The model and results presented here are a step towards a coherent model of interfirm competition network formation in particular, and dynamic perspective of economic geography in general. More research is needed in several directions. In this paper we consider the entry of new firms, their links, and dynamics of firm size as the only processes affecting the formation of the network. However, a more realistic description of the formation of competition networks should take into account the effect of various local events on the large topology of the network, including the formation of new competition links between existing firms, dissolution of existing competition links, shifting (or rewiring) of existing competition links, exit of existing firms, and merging of existing firms. The relative frequency of these local processes will determine (combined with geographic and population density effects) to a large degree the structure of competition networks. Moreover, other tie formation mechanisms operating at the microlevel could be considered such as similarity/dissimilarity in size, performance, or financial indicators between pairs of potential competing firms. While it is theoretically possible to incorporate the above modifications, the scarcity (at this point) of longitudinal competition and firm-specific data over significant time periods make it difficult to validate the model or parameterize it for simulation and prediction purposes.

The simple competition models introduced here offer an evolutionary perspective on economic geography and market structure that significantly extends traditional notions of economic competition and geographical clusters. Combined with proper models of competition dynamics, it also opens up a new range of empirical and analytic possibilities in realistically examining the effect of interfirm competition on firm performance, strategy dynamics, price and output changes, technology diffusion, the emergence of fast-growing geographic clusters (hot spots), and many other phenomena of industry dynamics (*e.g.,* "Red Queen" dynamics [56]). Finally, the model provides a framework to study the ability of competition networks to be resilient (robust) to firm and economic fluctuations.

## ACKNOWLEDGMENTS

The authors wish to thank the Editor/referees for the valuable comments. In particular, the insightful comments of one of the referees were helpful in developing the second competition model.

## REFERENCES

[1] Albert, R. and A.-L. Barabasi. 2002. *Rev. Mod. Phys.* 74 47-97.

[2] Amaral, L. A. N., A. Scala, M. Barth´l´my, and H. E. Stanley. 2000. *Proc. Nat. Ac. Sci USA* 97 11149-11152.

[3] Boccaletti, S., V. Latora, Y. Moreno, M. Chaves, and D.-U. Hwang. 2006 Complex networks: structure and dynamics. *Physics Reports,* 424 175308.

[4] Costa, L. da F., F. A. Rodrigues, G. Travieso and P. R. Villas Boas. 2007. *Advances in Physics,* 56(1) 167242.

[5] Braha D. and Bar-Yam Y. 2007. The statistical mechanics of complex product development: empirical and analytical results. *Management Science* 53(7) 1127-1145.

[6] Amaral, L. A. N. and B. Uzzi. 2007. Complex systems–a new paradigm for the integrative study of management, physical, and technological systems. *Management Science* 53(7) 10331035.

[7] Lee, E., J. Lee, and J. Lee. 2006. Reconsideration of the winner-take-all hypothesis: complex networks and local bias. *Management Science* 52(12) 1838-1848.

[8] Barabasi, A.-L. and R. Albert. 1999. Emergence of scaling in random networks. *Science* 286 509-512.

[9] Albert, R., H. Jeong, and A.-L. Barabasi. 2000. Error and attack tolerance of complex networks. *Nature* 406 378382.

[10] Cohen, R., K. Erez, D. ben-Avraham, and S. Havlin. 2000. *Phys. Rev. Lett.* 85 4626.

[11] Pastor-Satorras, R. and A. Vespignani. 2002. Epidemic dynamics in finite size scale-free networks. *Physical Review E* 65(3) 035108-035111.

[12] Hayashi, Y. 2006. A review of recent studies of geographical scale-free networks. *IPSJ Trans. Special Issue on Network Ecology* 47(3) 776-785.

[13] Waxman, B. 1988. Routing of multipoint connections. *IEEE J. Selec. Areas Commun.* 6 16171622.

[14] Yook, S.-H., H. Jeong, and A.-L. 2002. Barabasi. Modeling the internet's large-scale topology. *Proceedings of the National Academy of Sciences* 99(21) 1338213386.

[15] Lakhina, A., J. W. Byers, M. Crovella, and I. Matta. 2002. On the geographic location of internet resources. *Proceedings of the ACM SIGCOMM Internet Measurement Workshop, Marseilles.*

[16] Gastner, M. T. and M. E. J. Newman. 2006. The spatial structure of networks. *The European Physical Journal* B 49 247-252.

[17] Guimerà, R., S. Mossa, A. Turtschi, and L. A. N. Amaral. 2005. The worldwide air transportation network: anomalous centrality, community structure, and cities' global roles. *Proc. Natl. Acad. Sci. U. S. A.* 102 7794-7799.

[18] Porta, S., P. Crucitti, and V. Latora. 2006. The network analysis of urban streets: a dual approach. *Physica A: Statistical Mechanics and its Applications* 369(2) 853866.

[19] Lim, M., D. Braha, S. Wijesinghe, S. Tucker, and Y. Bar-Yam. 2007. Preferential detachment in broadcast signaling networks: connectivity and cost trade-off. *Europhysics Letters* 79 (5) 58005-1–58005-6.

[20] Albert, R., I. Albert, and G.L. Nakarado. 2004. Structural vulnerability of the North American power grid. *Physical Review E* 69(2) 25103.

[21] Lambiotte, R., V.D. Blondel, C. de Kerchove, E. Huens, C. Prieur, Z. Smoreda, and P. Van Dooren. 2008. Geographical dispersal of mobile communication networks. *Physica A* 387 53175325.

[22] Eguıluz, V. M., D. R. Chialvo, G. A. Cecchi, M. Baliki, and A. V. Apkarian. 2005. Scale-free brain functional networks. *Physical Review Letters* 94(1) 18102.

[23] Liben-Nowell, D., J. Novak, R. Kumar, P. Raghavan, and A. Tomkins. 2005. Geographic routing in social networks. *Proc. Natl. Acad. Sci. U. S. A.* 102(33) 11623-11628.

[24] Rozenfeld, A. F., R. Cohen, D. ben-Avraham, and S. Havlin. 2002. Scale-free networks on lattices. *Phys. Rev. Lett.* 89 218701-1–218701-4.

[25] Herrmann, C., M. Barthelemy, and P. Provero. 2003. Connectivity distribution of spatial networks. *Phys. Rev. E* 68 026128-1–026128-6.

[26] Manna, S. S. and P. Sen. 2002. Modulated scale-free network in the euclidean space. *Phys. Rev. E* 66 066114-1–066114-4.

[27] Barthelemy, M. 2003. Crossover from scale-free to spatial networks. *Europhys. Lett.* 63 915-921.

[28] Krugman, P. 1996. *The Self-organizing Economy.* Blackwell, Malden, MA.

[29] Grabher, G. 2006. Trading routes, bypasses, and risky intersections: mapping the travels of 'networks' between economic sociology and economic geography. *Progress in Human Geography* 30(2), 163-189.

[30] Podolny, J. M. 1994. Market uncertainty and the social character of economic exchange. *Administrative Science Quarterly* 39 458-483.

[31] Gulati, R. 1995. Social structure and alliance formation patterns: a longitudinal analysis. *Administrative Science Quarterly* 40 619-652.

[32] Lincoln, J., G. Michael, and A. Christina. 1996. Keiretsu networks and corporate performance in Japan. *American Sociological Review* 61 67-88.

[33] Gulati, R. and M. Gargiulo. 1998. Where do interorganizational networks come from? *American Journal of Sociology* 104(5) 1439-1493.

[34] Stuart, T. E. 1998. Network positions and propensities to collaborate: an investigation of strategic alliance formation in a high-technology industry. *Administrative Science Quarterly* 43 668-98.

[35] Podolny, J. M. 1993. A status-based model of market competition. *American Journal of Sociology* 98 829-872.

[36] Gulati, R. 1999. Network location and learning: the influence of network resources and firm capabilities on alliance formation. *Strategic Management Journal* 20 397-420.

[37] Powell, W. W., D. R. White, K. W. Koput, J. Owen-Smith. 2005. Network dynamics and field evolution: the growth of interorganizational collaboration in the life sciences. *American Journal of Sociology* 110 1132-1205.

[38] Sorenson, O. and T. E. Stuart. 2001. Syndication networks and the spatial distribution of venture capital investments. *The American Journal of Sociology* 106(6) 1546-1588.

[39] Owen-Smith, J. and W. W. Powell. 2004. Knowledge networks as channels and conduits: the effects of spillovers in the Boston biotechnology community. *Organization Science* 15 5-21.

[40] Fleming, L. and O. Sorenson. 2001. Technology as a complex adaptive system: evidence from patent data. *Research Policy* 30 1019-1039.

[41] Sorenson, O. 2003. Social networks and industrial geography. *Journal of Evolutionary Economics* 13 513-527.

[42] Google, Inc. http://www.google.com/intl/en/jobs/locations.html. Accessed 22 October 2009.

[43] Stouffer, S. A. 1940. Intervening opportunities: a theory relating mobility and distance. *American*

*Sociological Review* 5 845-867.

[44] Zipf, G. K. 1949. *Human Behavior and the Principle of Least Effort.* Addison-Wesley, Reading, MA.

[45] Isard, W. 1954. Location theory and trade theory: short-run analysis. *Quarterly Journal of Economics* 68 305- 322.

[46] Zipf, G. K. 1946. The P1P2/D Hypothesis: On the Intercity Movement of Persons. *American Sociological Review* 11 677-686.

[47] Granovetter, M. 1985. Economic action and social structure: the problem of embeddedness. *American Journal of Sociology* 91 481-510.

[48] Braha D. and Y. Bar-Yam. 2004. The Topology of large-scale engineering problem-solving networks. *Physical Review E* 69(1) 016113-1016113-7.

[49] Stanley, H. E., L. A. N. Amaral, S. V. Buldyrev, P. Gopikrishnan, V. Plerou, and M. A. Salinger. 2002. Self-organized complexity in economics and finance. *Proc. Nat. Ac. Sci USA* 99 25612565.

[50] Rothenberg, R. B. 1995. Commentary: sampling in social networks. *Connections* 18(1) 104110.

[51] Lee S. H., P-J Kim, and H. Jeong. 2006. Statistical properties of sampled networks. *Physical Review E* 73 016102-1 – 016102-7.

[52] Sole R. V. and J. Bascompte. 2006. *Self-Organization in Complex Ecosystems.* Princeton University Press, New Jersey.

[53] Hannan M. T. and G. R. Carroll. 1992. *Dynamics of Organizational Populations.* Oxford University Press, New York.

[54] Podolny J. M., T. E. Stuart and M. T. Hannan. 1996. Networks, Knowledge, and Niches: Competition in the Worldwide Semiconductor Industry, 1984-1991. *The American Journal of Sociology* 102(3) 659-689.

[55] Ruef M. 2000. The Emergence of Organizational Forms: A Community Ecology Approach. *The American Journal of Sociology* 106(3) 658-714.

[56] Barnett W. P. and M. T. Hansen. 1996. The Red Queen in Organizational Evolution. *Strategic Management Journal* 17, 139-157.

[57] Clauset, A., C. R. Shalizi and M. E. J. Newman. 2009. Power-law distributions in empirical data. *SIAM Review* 51, 661–703.

[58] Mitzenmacher M. 2004. A Brief History of Generative Models for Power Law and Lognormal Distributions. *Internet Mathematics* 1, 226-251.

[59] Laherrere, J. and D. Sornette. 1998. Stretched exponential distributions in Nature and Economy: Fat tails with characteristic scales. *European Physical Journal B* 2, 525–539.

[60] Stock, J. H. and M. W. Watson. 2003. *Introduction to Econometrics.* Addison-Wesley, Boston.

[61] Long, J. S. 1997. *Regression Models for Categorical and Limited Dependent Variables*. Sage Publications, London.

[62] Vuong, Q. H. 1989. Likelihood Ratio Tests for Model Selection and Non-Nested Hypotheses. *Econometrica* 57, 307-333.

[63] Boes, D. C., F. A. Graybill and A. M. Mood. 1974. *Introduction to the Theory of Statistics*. McGraw-Hill, New York.

[64] Gripsrud, G. and K. Gronhaug. 1985. Structure and Strategy in Grocery Retailing: A Sociometric Approach. *Journal of Industrial Economics* 33, 339-47.

[65] Katz, J. 2002. Get Me Headquarters!. *Regional Review Q4*, 9-19.

[66] Carroll, G. 1985. Concentration and Specialization: Dynamics of Niche Width in Population of Organizations. *American Journal of Sociology* 90, 1263-1283.

[67] Baum, J. and S. Mezias. 1992. Localized Competition and Organizational Failure in the Manhattan Hotel Industry, 1898-1990. Administrative Science Quarterly 37, 580-604.

[68] Porac, J., H. Thomas, F. Wilson, D. Paton, and A. Kanfer. 1995. Rivalry and the Industry Model of Scottish Knitwear Producers. *Administrative Science Quarterly* 40, 203-227.

[69] Baum, J. and Haveman, H. 1997. Love Thy Neighbor? Differentiation and Agglomeration in the Manhattan Hotel Industry, 1898-1990. *Administrative Science Quarterly* 42, 304-338.

[69] Ijiri, Y., Simon, H. 1977. *Skew Distributions and the Sizes of Business Firms*. North-Holland, Amsterdam.

[70] Rosen, S. 1981. The Economics of Superstars. *The American Economic Review* 71, 845-858.

[71] O. Malcai, O. Biham, S. Solomon. 1999. Power-law distributions and Lévy-stable intermittent fluctuations in stochastic systems of many autocatalytic elements. *Phys. Rev. E* 60, 1299-1303.

[72] Axtell, R. L. 2001. Zipf Distribution of U.S. Firm Sizes. *Science*, 1818-1820.

[73] Sutton, J., 1997. Gibrat's Legacy. *Journal of Economic Literature* 35 (1), 40–59.

[74] We note that this kind of correlation was also observed for the geographical deployment of Internet routers (see references [14, 15]).

# SUPPLEMENTARY MATERIAL

## APPENDIX A. SAMPLE OF MICROSOFT'S COMPETITORS

| Operating Segment | Products & Services | Competitors | Rationale |
|---|---|---|---|
| **Client** | Windows Vista, Windows XP, standard Windows operating systems | Apple, Canonical, Red Hat | Unix-based commercial software products |
| | Windows Vista, Windows XP, standard Windows operating systems | Hewlett-Packard, Intel | Linux-based operating systems, which are available without payment under a General Public License |
| | Internet Explorer | Apple, Google, Mozilla, Opera Software Company | Software that compete with the Internet Explorer Web browsing capabilities of Windows products |
| | Windows Vista, Windows XP, standard Windows operating systems | Google | Google Android mobile operating system may reduce consumer demand for traditional PCs as user and usage volumes on mobile devices are increasing around the world relative to the PC |
| **Server & Tools** | Server operating system products | Hewlett-Packard, IBM, Sun Microsystems | Vertically integrated computer manufacturers that offer their own versions of the Unix operating system preinstalled on server hardware |
| | Server operating system products | Novell, Red Hat | Companies that supply versions of the Linux operating system, benefiting from a large number of server hardware and compatible applications produced by many leading commercial and non-commercial computer manufacturers and software developers |
| | Enterprise-wide computing solutions | IBM, Oracle, Sun Microsystems | Companies focused on the Java 2 Platform Enterprise Edition (J2EE) |
| **Online Services** | Online advertising platform (e.g., Microsoft adCenter), online information offerings (e.g., Bing), personal communication services (e.g., MSN Hotmail Plus) | AOL, Google, Yahoo! | Organizations that provide ways of connecting advertisers with audiences through content and online offerings of all types |
| **Business** | Microsoft Office system products | Adobe, Apple, Corel, Google, IBM, Novell, Oracle, Red Hat, Zoho, 37Signals, AjaxWrite, gOffice, ShareOffice, Socialtext, ThinkFree, Sun Microsystems | Significant installed bases with Corel's (WordPerfect) and IBM's (SmartSuite) office productivity products. Apple's application software products may be distributed with its PCs and through mobile devices. The freely downloadable OpenOffice.org project (adapted by IBM, Novell, Red Hat, and Sun Microsystems), and Web-based offerings (such as 37Signals and ThinkFree) provide an alternative to MS Office system products |
| **Entertainment & Devices** | Xbox 360 console and games; Xbox Live, Zune | Nintendo, Sony, Apple | Competition on the basis of product innovation, quality and variety, timing of product releases, and effectiveness of distribution and marketing. The market is characterized by significant price competition |

TABLE A: A sample of Microsoft's products & services, competitors, and the rationale for competition, as identified by Microsoft in its filed 2009 10-K report to the U.S. Securities and Exchange Commission.

APPENDIX B. FITTING PARAMETRIC DISTRIBUTIONS

**B.1 Common parametric statistical distributions**

We have considered several common functional forms for the various distributions (in-degree, out-degree, and link length distributions) examined in this paper. Table B.1 summarizes the basic functional forms that will be useful [2, 57-59].

| Name | $p(x) = Cf(x)$ |
| --- | --- |
| Power law | $Cx^{-\gamma}$ |
| Power law with cutoff | $Cx^{-\gamma}e^{-\lambda x}$ |
| Exponential | $Ce^{-\lambda x}$ |
| Stretched exponential | $Cx^{\beta-1}e^{-\lambda x^{\beta}}$ |
| Log-normal | $C\dfrac{1}{x}e^{\left[-\dfrac{(\ln x - \mu)^2}{2\sigma^2}\right]}$ |

TABLE B.1: Definition of several common parametric statistical distributions [2, 57-59]. The normalization constant $C$ is determined from the requirement that $\int_{x_{\min}}^{\infty} Cf(x)dx = 1$, where $x_{\min}$ denotes the lower bound of the range of possible values that the random variable can attain. In this paper, we use the continuous approximations for the discrete degree distributions, which are particularly accurate for very large-size networks.

**B.2. Estimating the parameters of the parametric distributions**

Nonlinear least squares is a general method for estimating the unknown parameters of a probability density function. However, the nonlinear least squares estimators are inefficient, and generate significant systematic errors under relatively common conditions [57]. For this reason, we use in this paper an alternative estimation technique, the maximum likelihood method, for fitting the parameterized models in Table B.1 to observed data. The likelihood function is the joint probability of the data, treated as a function of the unknown coefficients. The maximum likelihood estimation (MLE) of the unknown parameters consists of the values of the parameters that maximize the likelihood function [60]. The ML estimators are consistent, asymptotically efficient, and asymptotically normally distributed [60]. The ML estimators of the various models' parameters are shown in Tables B.2-B.4.

| Model | Maximum likelihood estimation [95% confidence interval] | |
|---|---|---|
| Power law | $\gamma = 1.62$ [1.6093, 1.6332] | |
| Power law with cutoff | $\gamma = 0.98$ [0.9435, 1.0084] | $\lambda = 0.04$ [0.0384, 0.0440] |
| Exponential | $\lambda = 0.12$ [0.1221, 0.1268] | |
| Stretched exponential | $\beta = 0.53$ [0.5067, 0.5437] | $\lambda = 0.58$ [0.5410, 0.6197] |
| Log-normal | $\mu = 1.08$ [1.0773, 1.0780] | $\sigma = 1.39$ [1.3914, 1.3923] |

TABLE B.2: Fitting parametric distributions to the out-degree distribution in Figure 1A in the main text. The parametrized models are fitted to the observed data using the method of maximum likelihood.

| Model | Maximum likelihood estimation [95% confidence interval] | |
| --- | --- | --- |
| Power law | $\gamma = 1.50$ [1.4948, 1.5145] | |
| Power law with cutoff | $\gamma = 0$ [-0.0004, 0.0004] | $\lambda = 0.12$ [0.1227, 0.1274] |
| Exponential | $\lambda = 0.12$ [0.1227, 0.1274] | |
| Stretched exponential | $\beta = 1.43$ [1.4011, 1.4506] | $\lambda = 0.04$ [0.0373, 0.0428] |
| Log-normal | $\mu = 1.96$ [1.9481, 1.9758] | $\sigma = 0.71$ [0.7046, 0.7251] |

TABLE B.3: Fitting parametric distributions to the in-degree distribution in Figure 1B in the main text. The parametrized models are fitted to the observed data using the method of maximum likelihood.

| Model | Maximum likelihood estimation [95% confidence interval] | |
| --- | --- | --- |
| Power law | $\gamma = 0.93$ [0.9273, 0.9299] | |
| Power law with cutoff | $\gamma = 0.17$ [0.1684, 0.1816] | $\lambda = 0.0005$ [0.0005, 0.0005] |
| Exponential | $\lambda = 6.56 \times 10^{-4}$ [$6.523 \times 10^{-4}$, $6.607 \times 10^{-4}$] | |
| Stretched exponential | $\beta = 0.95$ [0.9414, 0.9517] | $\lambda = 0.0010$ [0.0010, 0.0010] |
| Log-normal | $\mu = 6.61$ [6.6018, 6.6240] | $\sigma = 1.74$ [1.7314, 1.7471] |

TABLE B.4: Fitting parametric distributions to the link length distribution in Figure 4 in the main text. The parametrized models are fitted to the observed data using the method of maximum likelihood.

# APPENDIX C. DIRECT COMPARISON OF PARAMETRIC DISTRIBUTIONS

After fitting the parametric distribution models to the data sets as specified in Appendix B, we have conducted several statistical tests that can help in comparing the fit of the different distribution models. In this appendix we present the statistical tools (and the results obtained) that have been used to assess the fit of a distribution model and compare it to other distributions.

**Likelihood value.** The basic measure of how well the maximum likelihood estimation procedure fits is the likelihood value, $L$. The likelihood value $L$ can be used to compare the fit of multiple distribution models. The distribution with the largest $L$ value is then the better fit. Without loss of generality, the values reported in the tables below are the log-likelihood values.

**Likelihood ratio tests.** We compare two competing distribution models by calculating the logarithm $R$ of the ratio of their two likelihoods [57, 61]. This is equivalent to calculating the difference in their log-likelihoods. The first case to consider is when the two distributions are **nested**; meaning that all of the parameters of the smaller model must also be in the bigger model. For example, the "power law" distribution is nested within the "power law with exponential cutoff" distribution (see Table B.1). For nested models, we use a chi-square test to evaluate the significance of the difference between the log-likelihood values. More specifically, we calculate the test statistic $\chi^2 = 2R = 2[\text{LL(bigger model)} - \text{LL(smaller model)}]$. It is known that this test statistic is distributed as chi-square with degrees of freedom that are the difference between degrees of freedom for the bigger and smaller models [61]. For **non-nested** hypotheses, we use a method proposed by Vuong [62]. This method computes a p-value that indicates whether the observed value of $R$ (with a positive or negative sign) is statistically significant, and is thus sufficiently far from zero. A small p-value indicates that it is unlikely that the observed value of $R$ has occurred by chance, and that one model could be said to be a better fit to the data. More specifically, the p-value of the above test is given by $p = \text{erfc}(\frac{|R|}{\sqrt{2n}\sigma})$, where the function erfc($x$) is the complementary error function, $n$ is the sample size, and σ is the standard deviation of $R$ (see [57] for a nice exposition of this method).

| Model | Maximized Log-Likelihood | LR |
|---|---|---|
| Power law | - 32381 | 20.42*** |
| Power law with cutoff | -31050 | 1.93** |
| Exponential | - 32408 | 10.42*** |
| **Stretched exponential** | **-31015** | -- |
| Log-normal | - 31126 | 6.98*** |

TABLE C.1: Direct comparison of the out-degree probability distribution models estimated in Table B.2. For each distribution model, we give the maximized log likelihood, and the likelihood ratios for the alternatives relative to the distribution model with the largest log-likelihood value ('stretched exponential'). In all cases, we use the likelihood ratio tests for non-nested hypotheses, and quote the normalized log likelihood ratio $R/\sqrt{2n}\,\sigma$. Based on the the p-values generated by the likelihood ratio tests, we denote whether the normalized log-likelihood ratios are statistically significant at the *5%, **1%, or ***0.1% level. The results provide support for the stretched exponential distribution over the alternatives.

| Model | Maximized Log-Likelihood | LR |
|---|---|---|
| Power law | -38469 | 55.14*** |
| Power law with cutoff | -32496 | 11.71*** |
| Exponential | -32496 | 11.71*** |
| **Stretched exponential** | **-31916** | -- |
| Log-normal | -32043 | 2.43*** |

TABLE C.2: Direct comparison of the in-degree probability distribution models estimated in Table B.3. Log-likelihood ratios are statistically significant at the *5%, **1%, or ***0.1% level. For a description of the columns, see Table C.1. The likelihood ratio tests provide statistical support for the stretched exponential distribution.

| Model | Maximized Log-Likelihood | LR |
|---|---|---|
| Power law | -889390 | 106520[***] |
| **Power law with cutoff** | **-782870** | -- |
| Exponential | -783850 | 980[***] |
| Stretched exponential | -783660 | 14.20[***] |
| Log-normal | -807980 | 119.56[***] |

TABLE C.3: Direct comparison of the link length probability distribution models estimated in Table B.4. For a description of the columns, see Table C.1. For non-nested alternatives, we report the normalized log likelihood ratio $R/\sqrt{2n}\,\sigma$, while for the nested hypotheses 'power law' and 'exponential' we report the test statistic $\chi^2 = 2R$ as described above. Log-likelihood test statistics are statistically significant at the [*]5%, [**]1%, or [***]0.1% level. The results provide support for the power law with cutoff distribution.

APPENDIX D. DIRECT COMPARISON OF COMPETITION PROBABILITY MODELS

Visual inspection of Figure 3 -- which shows the probability $p(v_i \to v_j | d(v_i, v_j) = l)$ that two companies separated by a distance $l$ are related by a competition link -- leads us to consider the alternative competition probability models presented in Table D.1.

| Name | $p(v_i \to v_j | d(v_i, v_j) = l)$ |
|---|---|
| Bernoulli | $C$ |
| Power law | $Cl^{-\beta}$ |
| Power law with cutoff | $Cl^{-\beta} e^{-\lambda l}$ |
| Exponential | $Ce^{-\lambda l}$ |

TABLE D.1: Definition of several competition probability models $p(v_i \to v_j | d(v_i, v_j) = l)$, where $v_i$ and $v_j$ denote two companies separated by a distance $l$.

**D.1. Estimating the parameters of the competition probability models**

Theoretically, the maximum likelihood method described in Appendix B.2 could be used for fitting the parameterized models in Table D.1 to the observed binary data (i.e., whether a competition link exists or not). Unfortunately, the large size competition network, and the related extremely large number of binary events (more than 115 million), precludes the direct use of the maximum likelihood method. To get estimators that are computationally feasible, we use nonlinear least-squares estimation as described in Figure 3. The least-squares estimators of the various models' parameters are shown in Table D.2.

| Model | Least-squares estimation of parameters | | | Mean squared error (MSE) |
|---|---|---|---|---|
| Bernoulli | $C = 0.00077$ | | | $1.23 \times 10^{-7}$ |
| Power law | $C = 0.008$ | $\beta = 0.33$ | | $2.6 \times 10^{-8}$ |
| Power law with cutoff | $C = 0.01$ | $\beta = 0.38$ | $\lambda = -0.0001$ | $2.51 \times 10^{-8}$ |
| Exponential | $C = 0.0003$ | $\lambda = 0.001$ | | $6.28 \times 10^{-8}$ |

TABLE D.2: Fitting parametric distributions to the competition probability in Figure 3. The parametrized models are fitted to the observed data using the method of nonlinear least-squares estimation.

**D.2. Comparing the competition probability models**

We have used two ways when comparing the overall fit of the competing models. The first way employs the method of least squares presented above. Specifically, we measure the overall model fit by the mean squared differences between the actual and predicted competition probability values as shown in the third column (labeled "MSE") of Table D.2. The value of the MSE, however, will not reliably indicate which model is the better fit. In order to make a reliable choice between the competing models, we use the log likelihood ratio test described in Appendix C. To this end, we specify the likelihood function as follows. Let $k = (v_i, v_j)$ denote two companies separated by a distance $l$, and define $y_k = 1$ or $y_k = 0$ depending whether the two companies are related by a competition link or not. Define $p_k$ as the probability of observing whatever value of $y_k$ was actually observed for a given observation [61]:

$$p_k = \begin{cases} \Pr(y_k = 1 | d(v_i, v_j) = l) & \text{if } y_k = 1 \text{ is observed} \\ 1 - \Pr(y_k = 1 | d(v_i, v_j) = l) & \text{if } y_k = 0 \text{ is observed} \end{cases}$$

$\Pr(y_k = 1 | d(v_i, v_j) = l)$ is calculated based on the predicted competition probability values of the nonlinear least-squares estimation as described in Figure 3. The likelihood function is thus

$$L = \prod_{i=1}^{n} p_k = \prod_{y=1} \Pr(y_k = 1 | d(v_i, v_j) = l) \prod_{y=0} 1 - \Pr(y_k = 1 | d(v_i, v_j) = l)$$

where the index for multiplication indicates that the product is taken over only those cases where $y_k = 1$ and $y_k = 0$, respectively [61]. As mentioned in Appendix D.1, full calculation of the likelihood function is a prodigious task ($n > 115 \times 10^6$). To alleviate the computational problem, we use Monte Carlo sampling. Specifically, we randomly sample a large number of events (ordered pairs of companies and their associated distance), and use the likelihood function as a way to compare the competition probability models presented in Table D.1. The results of this procedure are shown in Table D.3. The likelihood ratio tests show that both the "power law" and "power law with exponential cutoff" are a better fit than both the "Bernoulli" model ($p \leq 0.1\%$ and 0.1%, respectively) and the "exponential" model ($p \leq 1\%$ and 5%, respectively). Moreover, when comparing the "power law" and "power law with cutoff", the large $p$-value (0.9557) tells us that the difference in their log-likelihoods is not statistically significant, and that the test does not favor either model over the other. By the parsimony principle, we choose the 'power law' competition probability model with a smaller number of parameters.

| Model | Log-Likelihood | LR |
|---|---|---|
| Bernoulli | -6440.5 | 3.12[***] |
| **Power law** | **-6384.07** | -- |
| Power law with cutoff | -6384.14 | 0.039 ($p$-value=0.96) |
| Exponential | -6403.26 | 1.99[**] |

TABLE D.3: Direct comparison of the competition probability models estimated in Table D.2. Statistically significant *p*-values are highlighted in bold. The log-likelihood function is calculated using Monte Carlo sampling of 1 million events (pairs of companies). The LR test is described in Appendix C. Log-likelihood test statistics are statistically significant at the [*]5%, [**]1%, or [***]0.1% level. The results indicate that both the "power law" and "power law with exponential cutoff" are a better fit than both the "Bernoulli' and 'exponential" models, and that the "power law" is a plausible alternative to the "power law with exponential cutoff."

## APPENDIX E. COMPARISON OF COMPETITION NETWORK MODELS

In this appendix we describe the method used to calibrate and compare the different network growth simulation models described in Section V, and the results obtained.

The empirically observed competition network generates the out-degree and link-length distributions as shown in Figures 1 and 4 in the main text. The fit of a network growth model can thus be measured, separately, with respect to each data set. A network growth simulation model is defined in terms of the two continuously varying parameters $\alpha$ and $\beta$. Determining the exact best-fitted values of the parameters is computationally impractical. Instead, we have simulated a variety of network growth models with varying parameters of $\alpha$ and $\beta$, and have determined the values of $\alpha = 0.85$ and $\beta = -0.3$ that minimize the Kolmogorov–Smirnov (K-S) statistic [63], which is defined as the maximum distance between the cumulative distribution functions of the observed out-degree (or link-length) data and the fitted simulation model. Similar procedure was used for Competition Model 2 (see Section V.2).

More specifically, given the set of out-degree (or link length) observations $X_1, \ldots, X_n$ taken from the sampled competition network (see Figures 1 and 4 in the main text), an empirical distribution function $F_{\text{real}}(x)$ is computed. We also compute the empirical distribution function $F_{\text{simulation}}(x)$, given the set of out-degree (or link length) observations $Y_1, \ldots, Y_n$ taken from the simulation results of a network growth model. The Kolmogorov–Smirnov (K-S) statistic is then defined as $D = \max_x |F_{\text{real}}(x) - F_{\text{simulation}}(x)|$.

The results of this procedure are shown in Tables E.1 and E.2 below. The tables also show the results of the best-fitted parametric models considered in Appendices B and C. The results indicate that Competition Models 1 and 2 give comparable results that are a better fit than the alternatives. All in all, the attractiveness of the network growth models is derived from their mathematical simplicity, and their ability to offer explanatory mechanisms and insights on the interfirm competition phenomenon.

| Model | Kolmogorov–Smirnov (K-S) statistic |
|---|---|
| Stretched exponential (parametric model) | 0.09 |
| Null model ( $\alpha = 0, \beta = 0$ ) | 0.11 |
| **Competition Model 1** ( $\alpha = 0.85, \beta = -0.3$ ) | **0.03** |
| **Competition Model 2** ($T = 100, \lambda \sim U(0.9,1.1), \beta = -0.3, \nu = 0.75, c = 0.23$) | **0.03** |
| Linear Preferential Attachment ( $\alpha = 1, \beta = 0$ ) | 0.04 |
| Gravity I ( $\alpha = 0, \beta = -1$ ) | 0.09 |
| Gravity II ( $\alpha = 0, \beta = -2$ ) | 0.1 |

TABLE E.1: Comparison of network growth and parametric models. For each model, we give the Kolmogorov–Smirnov (K-S) statistic, which measures the distance between the cumulative distribution function of the observed out-degree data and the cumulative distribution function corresponding to the out-degree data generated by the underlying model. The results provide support for Competition Models 1 and 2 over the alternatives.

| Model | Kolmogorov–Smirnov (K-S) statistic |
|---|---|
| Power law with cutoff (parametric model) | 0.08 |
| Null model ( $\alpha = 0, \beta = 0$ ) | 0.28 |
| **Competition Model 1** ( $\alpha = 0.85, \beta = -0.3$ ) | **0.06** |
| **Competition Model 2** ($T = 100, \lambda \sim U(0.9,1.1), \beta = -0.3, \nu = 0.75, c = 0.23$) | **0.06** |
| Linear Preferential Attachment ( $\alpha = 1, \beta = 0$ ) | 0.14 |
| Gravity I ( $\alpha = 0, \beta = -1$ ) | 0.31 |
| Gravity II ( $\alpha = 0, \beta = -2$ ) | 0.71 |

TABLE E.2: Comparison of network growth and parametric models. For each model, we give the Kolmogorov–

Smirnov (K-S) statistic, which measures the distance between the cumulative distribution function of the observed link-length data and the cumulative distribution function corresponding to the link-length data generated by the underlying model. The results provide support for Competition Models 1 and 2 over the alternatives.